# Plasmonic nanopore to monitor in-pore chemistry

Weihong Wang[a], Shukun Weng[b], Ali Douaki[b,c], German Lanzavecchia[b,c], Yanqiu Zou[d], Qifei Ma[a], Huaizhou Jin[e], Roman Krahne[b], Shangzhong Jin[a*], Makusu Tsutsui[*f], and Denis Garoli[a,b,c*]

In solid-state nanopores, achieving reliable control over pore aperture opening and closing (gating) remains a major challenge. Gating can be driven by the applied voltage involving electrically tunable chemical reactions, achieved by selecting appropriate compounds within the nanopore volume. In particular, cyclic precipitation and dissolution of metal phosphates can be triggered by regulating cation transport through an applied transmembrane voltage, thereby enabling reversible pore gating. Under negative bias, metal phosphate precipitates form inside the pore, obstructing ion flow and reducing current. Switching the polarity dissolves the precipitates, restoring ionic conductance. This process effectively produces a nanofluidic diode characterized by a remarkably high rectification ratio. To probe these localized chemical reactions more directly, we employed a plasmonic nanopore that generates strong confined fields, enabling surface-enhanced Raman scattering (SERS) measurements within the nanopore volume during cyclic gating. These measurements not only validate the proposed in-pore chemistry but also highlight the potential of plasmonic nanopores as powerful tools for monitoring nanoscale chemical processes with high spatial resolution.

## Introduction

Over the past two decades, solid-state nanopore technology has been the focus of extensive research, particularly aimed at creating artificial systems that mimic biological nanopores—transmembrane proteins capable of regulating the passage of individual molecules and ions[1–4]. While the primary effort has traditionally centered on sequencing applications—initially for single DNA molecules and more recently for single proteins [5–9]—the scope of interest in nanopore systems has now expanded considerably. Among various research directions, the use of solid-state nanopores for constructing nanofluidic circuits—such as ionic transistors and diodes—has recently attracted significant attention [2,10–12]. To develop a nanofluidic ionic system that mimics the behavior of a diode, precise control over the nanopore's geometry and surface charge is essential. This is because ion selectivity, the key mechanism to be exploited, cannot be effectively achieved with conventional $Si_3N_4$- or $SiO_2$-based nanopores[13,14]. In this context, a novel strategy for fabricating solid-state nanopores exhibiting diode-like properties, characterized by exceptionally high ion current rectification (ICR)[15], has been recently demonstrated. This method relies on the so-called "in-pore chemistry," where two solutions interact at the interface between the Cis and Trans reservoirs of the nanopore chamber—an interface effectively defined by the nanopore volume itself. In particular, this can be possible by using a thin $Si_3N_4$ membrane including a nanopore connecting two distinct ionic reactant solutions for precise confinement of the ion electromigration. These ions are capable of inducing electrically tunable metal phosphate precipitation/dissolution in the pore. This can be applied to the preparation of advanced nanofluidic devices including diodes and memristors, where the use of easy-to-fabricate large pore size to achieve high performances fluidic diodes represents a significant plus with respect to the conventional nano-scale systems[16–18].

To improve our understanding on the demonstrated "in-pore chemistry", here we investigated the chemical reaction inside the nanopore using optical spectroscopy. Considering that the reaction takes place in a very small volume (with nanopore's diameter of 200 nm and thickness of 50 nm, the volume is of the order of $10^{-3}$ fL), the chemical reaction can be monitored only using a highly localized label free spectroscopy. In particular, Surface Enhanced Raman Spectroscopy (SERS) represents a valuable technique where the electromagnetic field can be squeezed in the nanopore volume thanks to a proper design of a plasmonic nanopore[4,19]. Therefore, in this work we replaced the $Si_3N_4$ nanopore that we used previously [15] with a plasmonic nanopore designed as a thin gold layer on a $Si_3N_4$ membrane where the nanopore is surrounded by a bulls-eye grating with a period matching the excitation of the surface plasmon polaritons at the laser wavelength used for the excitation. As extensively proved in literature[19–23], this structure enables a strong near-field enhancement inside the nanopore, therefore facilitating measurements of the vibrational spectra during the precipitation/dissolution reactions.

## Results and Discussion

The bulls-eye integrated plasmonic nanopores were fabricated by means of focused ion beam (FIB) milling in a 50 nm thick $Si_3N_4$ membrane supported on a Si wafer. An initial pore's diameter of about 200 nm was prepared[15]. The sample was then modified with an additional thin layer of Ti/Au (5/25 nm) and a successive FIB sculpturing was used to prepare a bull-eye grating with a grating period matching the surface plasmon polariton wavelength ($\lambda_{SPP}$) for an incident light at 532 nm[24,25]. Figure 1 shows an illustration of the process and reports SEM images of the fabricated sample.

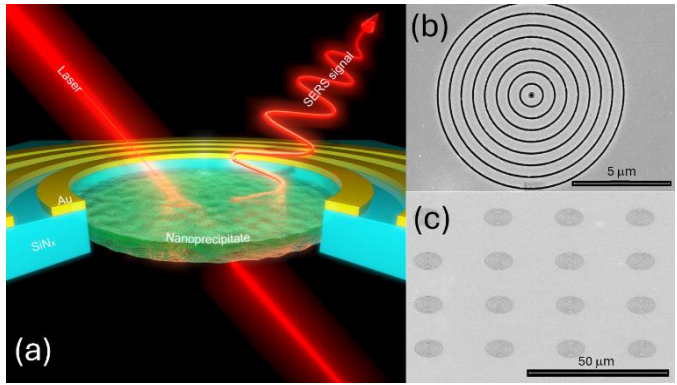

**Fig. 1.** Concept and fabrication. (a) Schematic illustration of the precipitation/dissolution at the nanopore aperture that can be monitores with SERS; (b) top view of a single plasmonic nanopore with a bulls-eye grating around it; (c) tilted view of an array of bulls-eye decorated nanopores.

As a first step we characterize the electrical properties of the nanopore to verify if our design comprising the Au layer results in the desired diode-like response. Here the Au layer can act as a floating electrode under the external applied field, and ion selectivity has been reported for gold nanopores[26]. Therefore, we first measured the current from the nanopore applying a transmembrane voltage ($V_b$) with a scan rate ($r_{scan}$) of 60 mVs$^{-1}$ in phosphate-buffered saline containing 1.37 M NaCl at pH 7.4 (hereafter, NaCl solution refers to the PBS unless otherwise denoted). In all our experiments we observed a perfect ohmic behavior with a slope ($G_{pore} = I_{ion}V_b^{-1}$) of 0.5 µS (Fig. 2), consistent with Maxwell–Hall's model for $d_{pore}$ of about 200 nm and solution resistivity ($\rho_{NaCl}$) of 0.08 Ωm. The same set of measurements have been performed replacing PBS respectively with CaCl$_2$ and MnCl$_2$ (concentration 2 mol/L) on both sides of the microfluidic chamber. These two solutions demonstrated to be very effective in the nanoprecipitation process that will be discussed later. As can be seen from Fig. 2, also in these cases, symmetric configurations on the Cis and Trans sides produce linear ohmic I-V response[27]. The nanopore's conductances where, 0.33 µS and 0.15 µS, respectively for CaCl$_2$ and MnCl$_2$.

A completely different response is expected when switching from a symmetric to asymmetric configurations of Cis and Trans sides of the chamber. In particular, using PBS at the Trans side and filling the Cis compartment with CaCl$_2$ or MnCl$_2$ salts (same concentrations used before) resulted in significant asymmetry in the $I_{ion}$–$V_b$ characteristics. As recently reported[15], CaCl$_2$ and MnCl$_2$ exhibited strong suppression of the ionic current at $V_b$<0. In the case of CaCl$_2$, for instance, the rectification ratio ICR = |$I_{max}/I_{min}$| reached over 40,000 (Fig. 2b), where $I_{max}$ and $I_{min}$ are $I_{ion}$ at $V_b$ = 0.6 V and -0.6 V, respectively. We note that relatively small $V_b$ values reach very high ICR values, and consequently with higher $V_b$ values the saturation of the current range that can be measured in the nanopore reader was reached (max I = 200 nA). These ICR values are more than an order of magnitude higher than the state-of-the-art systems based on nanofluidic diodes[28,29,30]. The rectification behavior depends on the voltage scan speed (Fig. 3c), reaching an ICR value of 20.000 (Fig. 2b) for a scan rate of 240mV/s with a very low and stable $I_{ion}$ at $V_b$= -0.6 V, corresponding to a nearly complete blockage of the translocation of high-concentration cations and anions under negative bias. In our previous study[15], we demonstrated that this ionic rectification cannot be justified considering concentration polarization or ion selectivity[29,30]. Electroosmotic flow also fails to explain the phenomenon as it works only under large transmembrane salinity difference [30,31], while we confirmed that we can detect only weak rectification by the ionic current measurements under salinity difference of the same types of salts at the cis and trans sides[15].

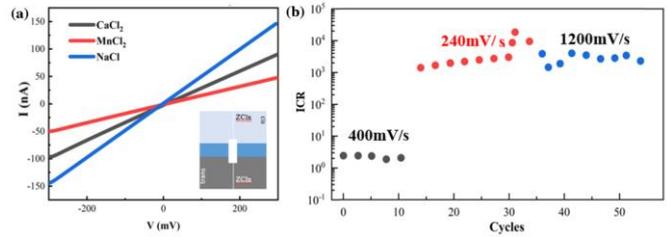

**Fig. 2.** (a) The ion transport properties under no salt difference (NaCl(blue), MnCl$_2$(red), and CaCl$_2$ (black)). The salinity was also the same at cis and trans(1.37 M for NaCl and 2 M for the rest of the salts). (b) ICR values obtained for CaCl$_2$ system at different scan rates.

We showed that such diode-like behaviour of the nanopore system is due to the formation of a solid precipitate via $V_b$-derived reactions that occluded the space for the electromigration of ions through the pores. Hence, we are observing precipitation reactions between divalent cations with the phosphoric acid in the electrolyte buffer to form metal phosphates. This can be understood considering the components in our chamber. The main ionic components of PBS in water are: Na$^+$, Cl$^-$, HPO$_4^{2-}$ and H$_2$PO$_4^-$ (the phosphate buffer pair). A divalent cloride compond in acqueous solution dissolves as : XCl$_2$ (s) → X$^{2+}$ (aq) + 2Cl$^-$ (aq), where X represents here Ca and Mn. Calcium and Manganese ions (X$^{2+}$) are known to react with phosphate ions to form insoluble phosphate salts, with a pH dependent reaction:

$$3X^{2+}+2PO_4^{3-} \rightarrow X_3(PO_4)_2 \qquad (1)$$

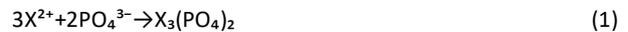

In PBS, most phosphate is present as HPO$_4^{2-}$ or H$_2$PO$_4^-$, but under physiological pH, there is still sufficient PO$_4^{3-}$ to trigger partial precipitation of salts phosphate (especially hydroxyapatite or amorphous X–P compounds). Therefore, the following reaction most likely happens at the interface between Cis and Trans sides (the nanopore's volume):

$$3XCl_2 + 2Na_3PO_4 \rightarrow X_3(PO_4)_2 + 6NaCl \qquad (2)$$

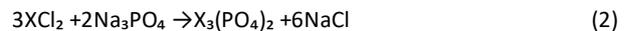

This reaction is strongly pH dependent as also demonstrated in our previour experiments[15], and therefore we use PBS at physiological pH.

Hence, the electrical response observed in our nanopores is ascribed to a solid-phase precipitate that clogs the aperture of the pore where the two solutions are mixed. The possibility to drive this reaction by external electrical potential is fundamental to provide the diode-like response and the significant ICR values reported.

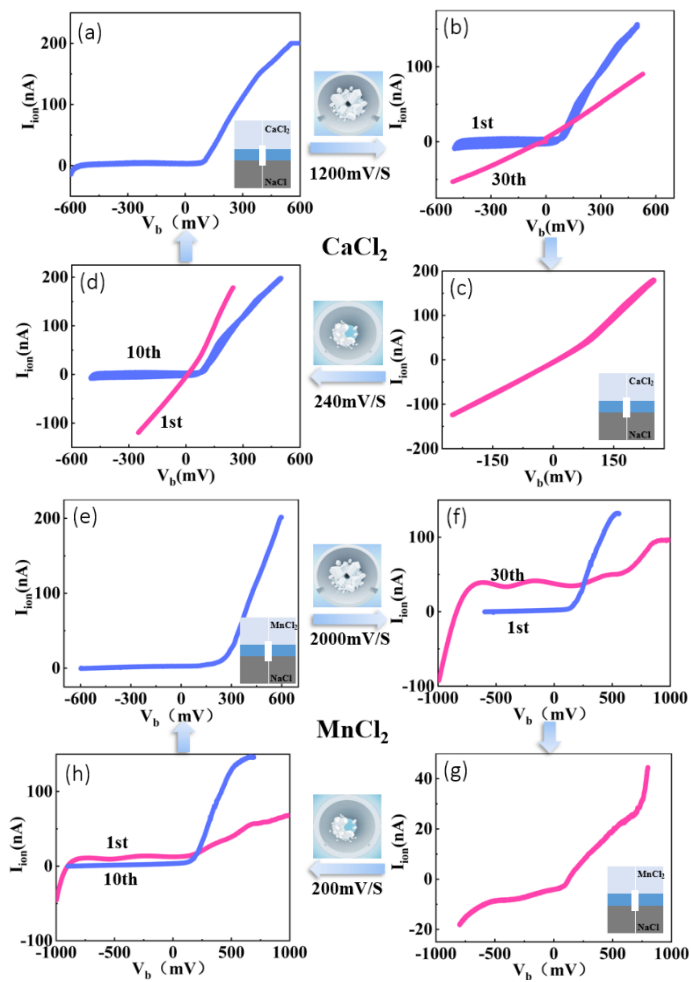

**Fig. 3. CaCl₂ system.** (a) In the CaCl$_2$/NaCl system, a 200 nm nanopore exhibits diode-like behavior at a voltage scan rate of 240 mV/s. (b-c) Repeated scanning at a faster rate (1200 mV/s) gradually opens the nanopore, leading to resistor-like behavior. (d) Subsequent scanning at 240 mV/s restores the diode characteristics of the nanopore. **MnCl₂ system.** (e) In the MnCl$_2$/NaCl system, the nanopore exhibits diode characteristics at a scan rate of 200 mV/s. (f-g) When the scan rate is increased to 2000 mV/s with repeated scanning, resistor-like behavior emerges. (h) Subsequently, scanning at 200 mV/s restores the diode characteristics of the nanopore.

Figure 4 reports the characterizations performed on the CaCl$_2$ system using the electron microscopy combined with energy dispersion spectroscopy (EDS). The morphological and compositional characterizations of the nanopore were done at the two stages corresponding to $I_{max}$ and $I_{min}$, i.e. when the external potential was $V_b$ = 0.6 V and −0.6 V, respectively. Figs. 4a and 4b clearly illustrate what happen inside the nanopore using the plasmonic configuration studied here. Additional tests using bare Si$_3$N$_4$ nanopores are reported in Supporting Information, confirming the precipitation and dissolution reactions. To note, these micrographs can be obtained stopping the reaction at the specific $V_b$ and resulted to be reproducible after multiple cycles. EDS maps revealed the presence of oxygen and phosphor inside the nanopore's cavity, as expected, even if the sentivity of the technique is probably not enough to obtain a convincing results. We then performed optical spectroscopic analyses by means of Raman spectroscopy. The goal was to obtain the spectrum from the nanopore's cavity in order to monitor the localized chemical reaction. Before to do it, the Raman spectra of the individual compounds were collected preparing them as dry droplet, in the case of PBS, and as dry powders, in the cases of CaCl$_2$ and MnCl$_2$. Fig. 5a reports the obtained vibration signatures where it is possible to observe how for PBS three peaks dominates, corresponding to 877 cm$^{-1}$, 979 cm$^{-1}$ and 1077 cm$^{-1}$, respectively. They are associated with the phosphate (PO$_4^{3-}$) buffer system, specifically the vibrations of the hydrogen phosphate (HPO$_4^{2-}$) and dihydrogen phosphate (H$_2$PO$_4^-$) ions [32]. In the cases of CaCl$_2$ and MnCl$_2$, on the contrary, only a main peak at around 1600 cm$^{-1}$ appeared. This can be associated with the H-O-H bending vibration of the water of crystallization [33]. In the nanopore system we are not dealing with salts powders, but we are interested to monitor the precipitation and dissolution processes driven by the external applied electrical potential. Therefore we also collected the Raman signatures of the precipitated salts as reported in eq. (1). To do that, we mixed the compounds using the same concentrations and pH conditions used in the nanopore experiments and after letting them dry we collected the spectra. In both the cases (Ca$_3$(PO$_4$)$_2$ and Mn$_3$(PO$_4$)$_2$), we can observe the presence of a main peak at around 960 cm$^{-1}$, clearly corresponding to the vibration of the phosphate group [34,35]. A minor broad peak at around 1600 cm$^{-1}$ can be also detected, as expected, to be associated with with the H-O-H bending [33,36].

Having a clear view on the vibrational signatures of the chemicals involved in our reactions, it is possible to proceed with the Raman experiments on the nanopore systems. Considering that this reactions have been firstly demonstrated in a Si$_3$N$_4$ pore, we performed the precipitation / dissolution processes focusing the laser over the nanopore directly prepared in a bare Si$_3$N$_4$ (the fabrication included both single nanopore and an array of nanopores in order to increase the reaction volume (see methods section)). Cycling them applying external electrical potential as discussed above and in our previous work [15] we can obtain the precipitation and dissolution processes.

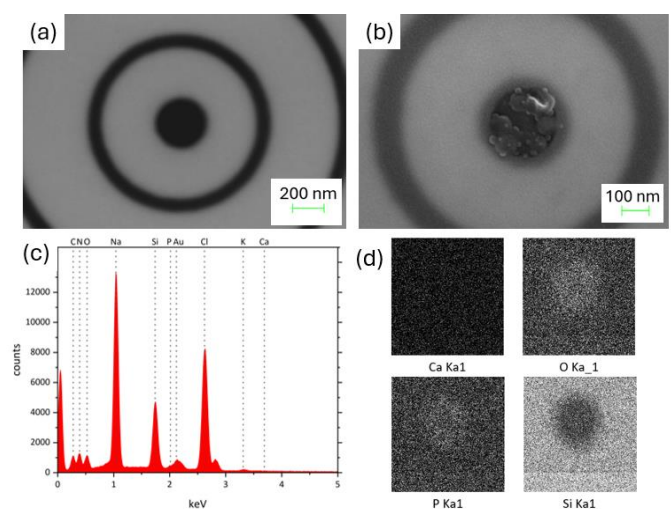

**Fig. 4.** SEM and EDS of the changes within the nanopore. (a) SEM micrograph of the bare plasmonic pore; (b) CaCl$_2$ precipitate inside the pore; (c) EDS spectrum; (d) EDS maps related to Ca, O, P and Si (additional data in the SI).

Considering the required integration time for the Raman acquisitions (tens of seconds), it was not possible to detect the dynamics of the precipation reaction. On the contrary, stopping the reaction at $V_b$ = 0.6 V and -0.6 V, we know that the nanopore(s) will be completely open and completely clogged, respectively. Important to note, the Raman measurements were done using a transparent microfluidic chip compatible with the simultaneous electrical and optical measurements, as demonstrated by Douaki et al. in Ref. [19]. This enables the monitoring of the reaction inside the nanopore. In order to avoid the background signal from the PBS solution in the Trans volume, we excited the pore from the Cis side where only the chloride salts are present. Testing both a single nanopore and an array of 16 pores (prepared in close proximity) it was not possible to detect convincing spectra to be ascribed to the precipitated salts. This because the volume of detection is spatially limited by the nanopore's diameter. While with the dry precipitates we can collect the signals from large area and large volume samples, here it has not been possible to obtain enough photons from the excited materials. This is the main motivation for the choice of the plasmonic nanopore. In fact, replicating the experiment replacing the bare Si$_3$N$_4$ pore with the single plasmonic nanopore (illustrated in Fig. 1b) it was easily possible to collect the SERS signatures from the nanopore in the two main configurations (open and clogged) using CaCl$_2$. As can be seen in Fig. 5c, there are clear differences in the Raman spectra associated with open and close pore. In particular, switching from the open to the close configuration, the peak at around 970 cm$^{-1}$ appears, confirming the formation of the phosphate precipitate. While the other spectral regions are not changed, we also observed and increased intensity for the peak at around 1600 cm$^{-1}$, associated with the H-O-H bending. Performing the same experiment by using MnCl$_2$, it resulted to be more challenging to collect the signatures from a single plasmonic nanopore. Therefore, in order to increase the detection volume, we reduced the laser spot size and we explored a configuration comprising 16 nanopores in array (Fig. 1c). In this case, it has been again possible to clearly detect the peak associated to the phosphate salts precipitation, presents only in the spectrum related to the clogged pore, while it always disappeared switching to the open pore configuration.

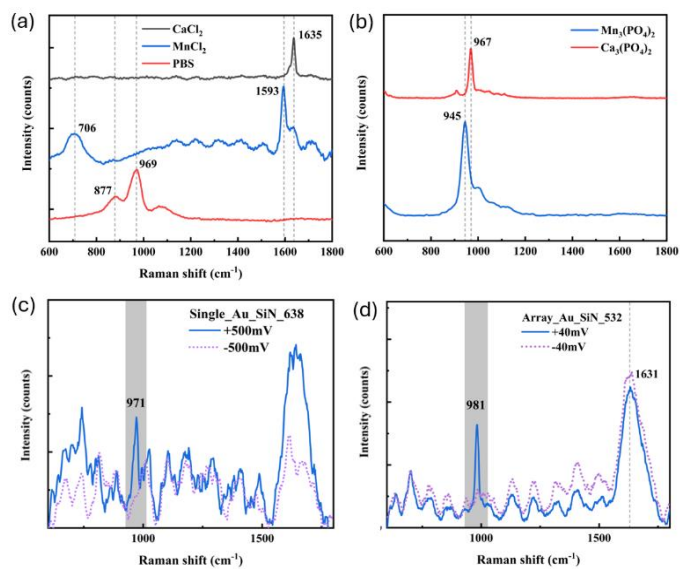

**Fig. 5.** (a) Raman spectra of powdered samples (PBS(red), MnCl$_2$(blue), and CaCl$_2$(black)). (b) Raman spectra of the precipitate (Ca$_3$(PO$_4$)$_2$ (red), Mn$_3$(PO$_4$)$_2$ (blue)). (c) Raman spectra of nanopore opening/closing in a Ca-based system (closed (blue), open (purple)). (d) Raman spectra of nanopore opening/closing in a Mn-based system (closed (blue), open (purple))

## Conclusions

In summary, this work demonstrates the advantages of plasmonic nanopore approach for probing nanoscale chemical processes. By confining intense electromagnetic fields to the attoliter-scale volume within a pore, the system achieves enhanced spatiotemporal resolution for monitoring localized chemical reactions via the ionic current. Simultaneously, SERS signals from the plasmonic hotspot provide molecular vibrational signatures that serve as rich chemical "fingerprints" of the reactants and products. This synergy between label-free electrical and optical readouts enables real-time in situ analysis of dynamic nanoscale chemical processes that are inaccessible to conventional analytical tools.

Beyond validating the specific in-pore precipitation/dissolution reaction, the strengths demonstrated here translate into broad practical applications across chemistry and biotechnology. In biosensing, a plasmonic nanopore can serve as a single-molecule analytical tool capable of identifying biomolecules such as proteins or nucleic acids by their vibrational spectra without the need for fluorescent or enzymatic labels, thereby enabling rapid diagnostics and sequencing. In catalytic reaction monitoring and synthetic chemistry, this platform allows researchers to observe chemical transformations in situ under realistic conditions, capturing transient intermediates and reaction kinetics as they occur. Such real-time insights into reaction pathways can inform catalyst design and process optimization and might even enable the discovery of novel reaction intermediates or products. By exploiting the nanopore as a nanoreactor with finely tunable conditions, one could potentially synthesize new compounds that are challenging to produce in bulk solution. Overall, this approach could transform how we study and control nanoscale chemistry, offering unprecedented spatiotemporal resolution, molecular specificity, and operando monitoring capabilities for next-generation biosensors, catalysts, and nanofluidic devices.

## Materials and methods

### Nanopore fabrication

Free standing SiN membrane chips were fabricated using a standard MEMS procesure based on commercial double-sided 100 nm LPCVD SiN-coated 500 μm silicon wafer. On on side, an array of ~850 μm SiN squres with 5mm periods was defined by UV photolithography (S1813), after RIE ething the SiN with $CHF_3$ and $O_2$, the wafer was cleaned by acetone and O2 plasma before soak into 32% KOH solution under 90 degree for few hours to to remove the silicon inside the UV lithography difined window thus make the other side SiN a suspended membrane. After sputter 5/30 nm Ti/Au on top of the SiN membrane, 18 pA focused ion beam (FIB, FEI 650) with 30 kV accelerate speed was used to drill ~ 200 nm pore and Au bullseye nanostructures around the pore.

### Ionic current measurements

Electrical measurements were conducted via a costum made microfluidic cell containing Ag/AgCl electrdoes then using the electrical reader from Elements srl. Voltage ramps were biased to one of the Ag/AgCl and the transmembrane ionic current was recorded through the other Ag/AgCl. Each measurement was carried out more than three times of Vb scans to obtain the arithmetic average of the ionic current, which was used for the evaluation of the in-pore reaction- mediated ionic current characteristics .

All salts were used as received without additional purification: NaCl (>99.5% purity), $CaCl_2$ (>95.0% purity), and $MnCl_2$ (>99.0% purity) were obtained from Sigma Aldrich.

To avoid cross-contamination and unwanted precipitation when changing between different electrolyte solutions, the following cleaning protocol was strictly followed: the existing solution was completely aspirated from the microfluidic cell with a pipette, after which both the inlet and outlet channels were thoroughly rinsed (at least five times) with Milli-Q water. Only then was the new salt solution introduced. This step is critical because direct contact between certain electrolyte combinations (particularly those containing phosphate buffers and divalent cations) can trigger immediate precipitation of insoluble salts, which would contaminate the nanopore walls, the membrane surface, and the entire fluidic system.

### SEM and EDS

SEM and EDS images were take from FEI 650 dual-beam system under 5 kV and 2.4 nA configuration. The chips with open or close state was generated by the above mentioned electrical measurements, after applying positive and negative potential for few seconds to form the precipitation inside the pore or resolve it, the chips then were take out from the measurements and dry in the air before mount into the dual-beam system for the secondary electron and EDS image.

### In-Pore Raman Spectroscopy

SERS measurements were performed using a HORIBA XploRA™ PLUS Raman spectrometer (Horiba Jobin Yvon, Kyoto, Japan) equipped with 50X long-focal-length objective (NA = 0.50). Visible Raman spectra of in-pore chemistry were measured using 532 nm laser at 10% power and spectra were acquired with a 600 gr/mm grating, 270 s exposure time, and one accumulation. The same laser and acquisition parameters were maintained for each measurement scan. During real-time SERS experiments, simultaneous electrical readout of the transmembrane ionic current was performed using the Elements srl reader described above.

## Acknowledgements

We are grateful for continued financial support from National Key Research and Development Project of China (No. 2023YFF0613603), National Natural Science Foundation of China (No. 22202167), Provincial Science and Technology Plan Project: Micro and Nano Preparation and Photoelectronic Detection (No. 03014/226063). D.G. thank the the HORIZON-MSCADN-2022: DYNAMO, grant Agreement 101072818. A part of this work was supported by the Japan Society for the Promotion of Science (JSPS) KAKENHI Grant Numbers 22H01926, 22H01410, and 24K21715. M. Tsutsui acknowledges support from Kansai Research Foundation for Technology Promotion.